\documentclass{article}
\usepackage{graphicx} % Required for inserting images
\usepackage[margin = 1in]{geometry}
\usepackage{amsmath,amsfonts}
\usepackage{algorithmic}
\usepackage{algorithm}
\usepackage{array}
\usepackage[caption=false,font=normalsize,labelfont=sf,textfont=sf]{subfig}
\usepackage{textcomp}
\usepackage{stfloats}
\usepackage{url}
\usepackage{verbatim}
\usepackage{graphicx}
\usepackage{cite}
\usepackage{hyperref}
\usepackage{booktabs}
\usepackage{natbib}
\usepackage{authblk}

\def\BibTeX{{\rm B\kern-.05em{\sc i\kern-.025em b}\kern-.08em
    T\kern-.1667em\lower.7ex\hbox{E}\kern-.125emX}}
\usepackage{natbib}
\title{Bot Identification in Social Media}
\author[1]{Dhrubajyoti Ghosh}
\author[2]{William Boettcher}
\author[3]{Rob Johnston}
\author[4]{Soumendra Lahiri}

\affil[1]{Department of Biostatistics and Bioinformatics, Duke University, Durham, NC, USA\\
\texttt{dhrubajyoti.ghosh@duke.edu}}

\affil[2]{Humanities and Social Sciences, North Carolina State University, Raleigh, NC, USA\\
\texttt{william\_boettcher@ncsu.edu}}

\affil[3]{School of Public and International Affairs, North Carolina State University, Raleigh, NC, USA\\
\texttt{bejohnst@ncsu.edu}}

\affil[4]{Department of Mathematics and Statistics, Washington University in St. Louis, St. Louis, MO, USA\\
\texttt{s.lahiri@wustl.edu}}
\date{}

\begin{document}

\maketitle

\begin{abstract}
Escalating proliferation of inorganic accounts, commonly known as bots, within the digital ecosystem represents an ongoing and multifaceted challenge to online security, trustworthiness, and user experience. These bots,  often employed for the dissemination of malicious propaganda and manipulation of public opinion, wield significant influence in social media spheres with far-reaching implications for electoral processes, political
campaigns and 
international conflicts. Swift and accurate identification of inorganic accounts is of paramount importance in mitigating their detrimental effects. This research paper focuses on the identification of such accounts and explores various effective methods for their detection through machine learning techniques. In response to the pervasive presence of bots in the contemporary digital landscape, this study extracts temporal and semantic features from tweet behaviors and proposes a bot detection algorithm utilizing fundamental machine learning approaches, including Support Vector Machines (SVM) and k-means clustering. Furthermore, the research 
ranks the importance of these extracted features for each detection technique 
and also 
provides uncertainty quantification using a distribution free method, called the conformal prediction, 
thereby contributing to the development of effective strategies for combating the prevalence of inorganic accounts in social media platforms.
\end{abstract}

\section{Introduction\label{sec:intro}}

Social media has emerged as a transformative force in contemporary society, reshaping the way individuals communicate, share information, and interact with the world. Over the past two decades, platforms like Facebook, Twitter, Instagram, and TikTok have permeated nearly every aspect of our lives, from personal relationships to political discourse and business marketing. According to a report by Pew Research Center, as of 2021, 69\% of adults in the United States use social media to connect with friends and family, emphasizing the central role these platforms play in interpersonal relationships \cite{pew}. Moreover, the increased use of visual content, such as images and videos, has revolutionized the way individuals express themselves, share experiences, and narrate their lives online \cite{duggan2013demographics}. 
Social media's influence extends beyond personal communication, as it has become an influential platform for political discourse and activism. The Arab Spring, the Occupy Wall Street, and the Black Lives Matter movement are just a few examples of how social media has facilitated the mobilization of social and political movements \cite{tufekci2017twitter}. The spread of information, the organization of protests, and the amplification of marginalized voices have all been made more accessible through social media platforms. However, the same platforms have also been criticized for facilitating the spread of misinformation, echo chambers, and polarization in political discourse \cite{pennycook2020implied}. Businesses and marketers have also recognized the potential of social media as a powerful tool for brand promotion and engagement with customers. Platforms like Instagram and Pinterest have given rise to influencer marketing, where individuals with large followings endorse products and services to their audience \cite{tanwar2021influencer}. Additionally, social media analytics provide valuable insights into consumer behavior and preferences, allowing companies to tailor their marketing strategies more effectively.

% The rise of social media has been one of the most important events in this century. Social media has given the common people a platform to share their views with the public. Information is now readily available to every corner of the world. Not long after its inception, social media started to prove effective in business, politics, society, and so on. Political campaign nowadays all over the world are heavily dependent on social media. The US presidential campaign of Barrack Obama, for example, extensively used different social media platforms, like Twitter , Facebook, Myspace, etc.

The proliferation of inorganic accounts, also known as bots, in social media has had a substantial and multifaceted impact on the digital landscape. These automated software programs, designed to mimic human behavior and engage with users on social media platforms, have influenced various aspects of online discourse. Bots have been instrumental in disseminating information rapidly, as demonstrated during major global events such as elections or crises, where they can amplify specific narratives or spread disinformation \cite{ferrara2016rise}. Moreover, their ability to generate high volumes of content and engagement can artificially inflate the apparent popularity of certain topics, products, or individuals \cite{bessi2016social}. While bots can contribute to the overall user experience by automating customer service and providing real-time information, they also pose significant challenges in terms of authenticity, credibility, and the potential to manipulate public opinion. As researchers continue to investigate the dynamics of bot-driven social media activity, it becomes increasingly critical to develop effective countermeasures to mitigate their negative effects and ensure the integrity of online interactions.
% Inorganic accounts, also known as bots, are an integral aspect of social media. Many companies and organizations uses bots to enable smooth functioning and better customer services. However, inorganic accounts can also have negative consequences. They can be used to spread fake rumors, or to publicize malicious propaganda. Undisclosed bots are also a security threat to organizations as well, which can cause serious monetary and publicity losses. 
In a social network, inorganic accounts controlled by a group can create and manipulate public opinions, and can play a large role in significant events, like elections or military conflicts \cite{ghosh2025thanos}. According to recent findings, about 19\% of Twitter traffic is created by bots \citep{bessi2016social}. Inorganic accounts or bots in social media has been extensively studied \citep{ferrara2016rise, subrahmanian2016darpa}. There has been multiple attempts in literature to identify inorganic social media accounts in order to stop malicious propaganda campaign as soon as possible. The BotOrNot \citep{davis2016botornot} software was made public in 2014 and uses several features from a Twitter user's tweet history in order to classify them as bots or non-bots. \citep{sayyadiharikandeh2020detection} used Random Forest Classifier on thousands of extracted features, \citep{mazza2019rtbust} proposed RTBust in order to use temporal distribution of retweets for unsupervised bot classification, \citep{cresci2017social} introduced the concept of social fingerprinting for bot detection techniques. However, bot detection remains an unsolved problem \citep{lee2011seven}, with plethora of research still going on in order to perfect social bot detection techniques.
% Bots can play a big role to magnify some topics, while downplaying other voices, thereby influencing people to a considerable extent. 

Inorganic accounts frequently exploit the principle of reciprocity, a phenomenon commonly observed in social media ecosystems. Typically, these automated entities amass a substantial following by employing strategies such as the follow-refollow approach. Subsequently, they engage in the dissemination of propagandistic content, which is then reshared or retweeted by genuine users who assume its credibility based on the perceived legitimacy stemming from the substantial follower count. Additionally, there exist other inorganic accounts that initiate the process of retweeting. Once a post garners significant shares or retweets, social media platforms are inclined to feature it prominently in the timelines of authentic users, who, in turn, propagate the content, resulting in an exponential increase in the original tweet's reach. Notably, bots that disseminate misinformation or attempt to influence the political beliefs of individuals can yield deleterious repercussions for society. This research article addresses the issue of bot detection through some existing machine-learning approaches. We primarily develop a static time bot detection algorithm, wherein we leverage the historical tweet data of users and employ machine learning algorithms to analyze extracted features, thereby classifying accounts as either organic or inorganic.

The paper is organized as follows: Section \ref{sec:methods} provides a detailed description of feature extraction from the tweet dataset, Section \ref{sec:dataDesc} provides a description of the dataset used for this study, Section \ref{sec:results} gives a detailed description of the result of bot identification, Section \ref{sec:uncertainty} provides a split conformal prediction analysis and Section \ref{sec:conslusion} provides a summary of the results in the paper.

\section{Methods\label{sec:methods}}

Let us denote the tweet history of an user $i$ as $\mathcal{V}_i = \{\mathcal{X}_i, \mathcal{T}_i\}$, where $\mathcal{X}_i = \{x_{i1}, x_{i2}, \ldots, x_{iJ_i}\}$ are the actual tweets of user $i$ and $\mathcal{T}_i = \{t_{i1, \ldots, t_{iJ_i}}\}$ are the corresponding times of the tweets. Suppose we have N users with $\{J_i\}_{i=1}^N$ tweets, then we have a dataset $\{\mathcal{V}_{i}\}_{i=1}^N$ of $N$ users, where $\mathcal{V}_i = \{\mathcal{X}_i, \mathcal{T}_i\}$, $\mathcal{X}_i = \{x_{ij}\}_{j=1}^{J_i}$ and $\mathcal{T}_i = \{t_{ij}\}_{j=1}^{J_i}$. Furthermore, we have two types of users, organic (regular human) and inorganic accounts or bots. Suppose we have $n_x$ organic users and $n_y$ inorganic users, then we will denote the organic dataset as $\{\mathcal{V}_i^{(O)}\}_{i=1}^{n_x}$ and the inorganic dataset as $\{\mathcal{V}^{(IO)}\}_{i=1}^{n_y}$, although the organic/inorganic labels
are typically unknown. 
We can extract two types of features from the dataset -- temporal features and semantic features. We will now briefly discuss the features extracted from the data for further analysis.

\subsection{Temporal Features\label{sec:temp}}

Given the (full) dataset \( \{\mathcal{V}_{i}\}_{i=1}^N \) with tweet histories \(\mathcal{V}_i = \{\mathcal{X}_i, \mathcal{T}_i\}\), where \(\mathcal{T}_i = \{t_{i1}, t_{i2}, \ldots, t_{iJ_i}\}\) represents the tweet timestamps for each user \(i\), we extract the following temporal features:

\subsubsection{Periodicity}

Periodicity identifies recurring patterns in tweet activity. For each user's tweet timestamps \(\mathcal{T}_i\), we calculate the periodicity by examining the power spectrum of the tweet frequency over time.
Let us define \(f_i(\tau)\) as the frequency of tweets over time \(\tau\) for user \(i\). The Discrete Fourier Transform (DFT) of \(f_i(\tau)\) provides the power spectrum:
\[
\mathcal{P}_i(\omega) = \left|\sum_{\tau} f_i(\tau) e^{-2\pi i \omega \tau}\right|^2,
\]
where \(\omega\) is the frequency variable. The primary periodicity \(\omega_i^*\) is then given by:
\[
\omega_i^* = \arg\max_{\omega} \mathcal{P}_i(\omega),
\]
which represents the frequency with the highest power, indicating regular tweet intervals (e.g., daily or hourly cycles). For our analysis, we have taken the time window $\tau$ as 3 hours, implying that a period of $8$ would mean a daily periodicity of tweet times.

\subsubsection{ARIMA-Based Features}
To capture the underlying temporal dependencies, we model each user's tweet time series \(\mathcal{T}_i\) using an Autoregressive Integrated Moving Average (ARIMA) process. This model helps extract parameters that describe statistical properties of tweet timing patterns. Let the ARIMA model for user \(i\) be given by:
\begin{align*}
 Y_{t_i} &= \phi_1 Y_{t_i-1} + \phi_2 Y_{t_i-2} + \ldots + \phi_p Y_{t_i-p} \\
 & \quad+ \theta_1 \epsilon_{t_i-1} + \theta_2 \epsilon_{t_i-2} + \ldots + \theta_q \epsilon_{t_i-q} + \epsilon_{t_i},   
\end{align*}

where \(Y_{t_i}\) represents the time series of 
\underline{tweet intervals}, \(\phi_k\) are autoregressive coefficients, \(\theta_k\) are moving average coefficients, \(\epsilon_{t_i}\) denotes white noise, and $p$ and $q$ are AR and MA orders respectively.

From the ARIMA model, we extract the following features:
\begin{itemize}
    \item \textbf{Log-Likelihood} \( L_i \): The log-likelihood value of the ARIMA fit, representing the quality of fit.
    \item \textbf{Sum of Squared Coefficients} \( \sum_{k} \phi_k^2 \): A measure of the impact of the autoregressive component.
    \item \textbf{Error Variance} \( \sigma^2 \): Variance of residuals \(\epsilon_{t_i}\), indicating the amount of noise in the model.
    \item \textbf{Fit Length}: The number of terms in the fitted ARIMA model, representing the complexity of temporal dependencies.
\end{itemize}

\subsubsection{Periodogram Shape (Local Maxima)}
To capture additional periodic characteristics beyond the primary periodicity, we examine the shape of the periodogram and the distribution of local maxima. 
% The periodogram of tweet timestamps \(\mathcal{T}_i\) is given by:
% \[
% P_i(\omega) = \frac{1}{J_i} \left|\sum_{j=1}^{J_i} t_{ij} e^{-2\pi i \omega j}\right|^2.
% \]
We identify local maxima \(M_i = \{\omega_{i1}, \omega_{i2}, \ldots, \omega_{ik}\}\) in the periodogram \(P_i(\omega)\), representing dominant recurring cycles beyond the main periodicity \(\omega_i^*\). This set \(M_i\) helps capture additional periodic patterns which are often characteristic of bots (regular patterns) versus human users (irregular patterns).

\subsection{Semantic Features\label{sec:semantic}}

Given each user’s tweet history \( \mathcal{X}_i = \{x_{i1}, x_{i2}, \ldots, x_{iJ_i}\} \), where \( x_{ij} \) represents the content of the \( j \)-th tweet by user \( i \), we define the following semantic features to capture patterns in linguistic and content usage.

\subsubsection{Lexical Diversity}
Lexical diversity quantifies the variety of vocabulary used by a user, helping to distinguish between users who use repetitive language (often characteristic of bots) and those with a more varied lexicon (typically human).
For each user \( i \), let \( V_i \) denote the set of unique words used across all tweets in \( \mathcal{X}_i \), and let \( W_i \) represent the total word count in \( \mathcal{X}_i \). Lexical diversity \( L_i \) is defined as:
\[
L_i = \frac{|V_i|}{W_i},
\]
where \( |V_i| \) is the number of unique words, and \( W_i = \sum_{j=1}^{J_i} |x_{ij}| \) represents the total word count across all tweets.

\subsubsection{Average Number of Words per Tweet and Variance}
The average number of words per tweet provides insight into tweet length and verbosity, while the variance indicates consistency or variability in tweet length.
For each user \( i \), the average number of words per tweet \( \mu_{W_i} \) is defined as:
\[
\mu_{W_i} = \frac{1}{J_i} \sum_{j=1}^{J_i} |x_{ij}|,
\]
where \( |x_{ij}| \) is the word count in the \( j \)-th tweet. The variance of the number of words per tweet \( \sigma_{W_i}^2 \) is given by:
\[
\sigma_{W_i}^2 = \frac{1}{J_i} \sum_{j=1}^{J_i} \left(|x_{ij}| - \mu_{W_i}\right)^2.
\]

\subsubsection{Relative Hashtag Frequency}
The frequency of hashtags used by a user can help differentiate between bots (which may overuse specific hashtags to promote content) and human users (who generally exhibit more varied usage).
Let \( H_{ij} \) be the number of hashtags in the \( j \)-th tweet of user \( i \). The relative hashtag frequency \( \eta_{H_i} \) for user \( i \) is defined as:
\[
\eta_{H_i} = \frac{1}{J_i} \sum_{j=1}^{J_i} H_{ij}.
\]
This feature provides the average number of hashtags per tweet for each user.

\subsubsection{Relative Frequency of Most Used Words}
This feature captures the concentration of certain words in a user’s tweets, with higher concentrations suggesting repetitive content typical of bots.
Let \( F_{i,k} \) denote the frequency of the \( k \)-th most used word by user \( i \) across all tweets in \( \mathcal{X}_i \). The relative frequency \( \rho_{i,k} \) of the \( k \)-th most used word is given by:
\[
\rho_{i,k} = \frac{F_{i,k}}{W_i},
\]
where \( W_i \) is the total word count across all tweets, as defined previously. Inorganic users tend to employ a limited set of words repeatedly, while organic users typically exhibit greater lexical diversity.. For practical analysis, we typically focus on the top few words (e.g., top 5) most frequently used by each user.

\subsubsection{Sentiment Score}
The sentiment score provides insight into the emotional tone of tweets, which can vary significantly between organic and inorganic users.
For each tweet \( x_{ij} \) of user \( i \), let \( S_{ij} \) denote the sentiment score derived from a sentiment analysis tool, such as AFINN, BING, or NRC lexicons. The average sentiment score \( S_i \) for user \( i \) is defined as:
\[
S_i = \frac{1}{J_i} \sum_{j=1}^{J_i} S_{ij}.
\]
This score helps assess whether users tend to post content with consistent emotional tones (often seen with bots) or varied tones (typical of human users).

\subsubsection{Summary of Semantic Features}
For each user \( i \), we represent the semantic features as a vector \( F_{sem}^i \):
\[
F_{sem}^i = \left( L_i, \mu_{W_i}, \sigma_{W_i}^2, \eta_{H_i}, \rho_{i,k}, S_i \right),
\]
where \( L_i \) is lexical diversity, \( \mu_{W_i} \) and \( \sigma_{W_i}^2 \) are the mean and variance of words per tweet, \( \eta_{H_i} \) is the relative hashtag frequency, \( \rho_{i,k} \) is the relative frequency of the most used words, and \( S_i \) is the average sentiment score.

\section{Data Description}
\label{sec:dataDesc}
In this study, we utilized a dataset obtained from the Bot Repository available at Botometer, as detailed in \cite{cresci2015fame}. The Bot Repository serves as a valuable resource for researchers, providing a comprehensive list of both identified organic and inorganic users. To compile this dataset, we leveraged the Twitter platform and employed the R programming language for data scraping. The dataset under analysis consists of the tweet histories of 470 verified organic users and 373 verified inorganic users, making it a substantial and diverse source for our research. Notably, the average number of tweets for verified organic users was calculated to be 3121.47, whereas the average for inorganic users stood at 2598.21 tweets. This dataset serves as a robust foundation for our analysis, facilitating the exploration of various machine learning techniques for the detection of inorganic accounts in social media platforms.

\begin{figure}[htbp]
\centerline{\includegraphics[width = 0.45\textwidth]{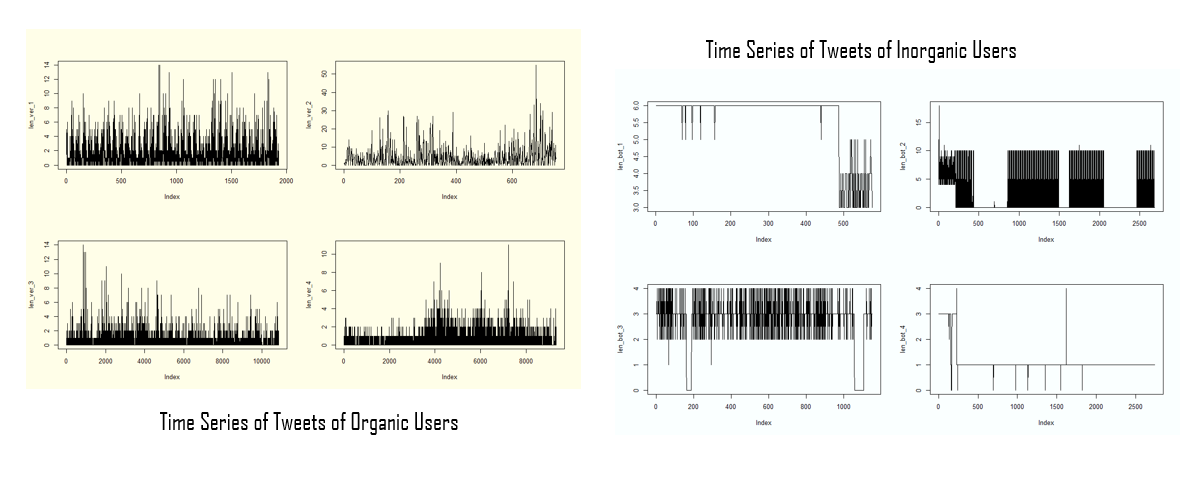}}
\caption{Time Series of Tweet History of Organic and Inorganic Users}
\label{fig1}
\end{figure}

% Figure \ref{fig1} gives the time series of tweet history of selected organic and inorganic users, using a granularity of three hours. The above sample demonstrates that the temporal activity of inorganic users has significant differences from that of organic users. One noticeable difference is that the inorganic users tends to be active uniformly for some fixed periods in the day, while the activity of organic users are fairly random. Hence, it is of importance to take temporal features into our classification algorithm. Following time series features were included in the analysis:

Figure \ref{fig1} presents the time series of tweet histories for a selection of both organic and inorganic users, capturing their activity patterns at a granularity of three hours. This illustrative sample underscores the substantial disparities in temporal activity between the two categories of users. Notably, inorganic users exhibit a characteristic pattern of uniform and consistent activity during specific fixed periods throughout the day, while the activity of organic users appears relatively random. Recognizing these temporal distinctions is imperative for our classification algorithm's effectiveness. Consequently, our analysis incorporates a set of time series features that encapsulate these patterns. The following temporal features were integrated into our analysis:

\begin{itemize}
    \item  A fundamental property of time series data is periodicity, which indicates how frequently observations are spaced in time. In our specific case, a periodicity of 8 corresponds to a 24-hour period, or one day. Our analysis revealed that most organic users exhibited periodicities of either 8 or 4, indicating daily or 12-hour recurring activity patterns. In stark contrast, inorganic users displayed significantly lower periodicity values, underscoring their distinct and irregular activity patterns. This characteristic is depicted graphically in the left panel of Figure \ref{fig2}, which illustrates the periodicity of tweet activity for both organic and inorganic accounts.
    % Periodicity of the Time series of tweet activity. A fundamental characteristic of time series is how frequently the observations are spaced in time, which is demonstrated by the periodicity property. For example, in this particular case, a periodicity of 8 would indicate a period of 24 hours, i.e. one day. As we can see for most organic users, the periodicity was 8 or 4, indicating a daily or 12 hour recurring activity. However, for inorganic users the periodicity was quite low, indicating that the activity pattern has a drastic difference from the inorganic users.

    \begin{figure}[htbp]
    % \centerline{
    \begin{center}
    \includegraphics[width = 0.45\textwidth]{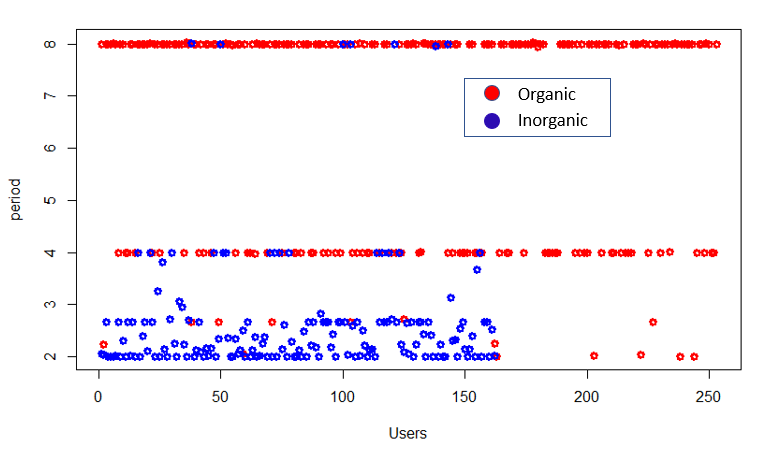}
    % }
    % \centerline{
    \includegraphics[width = 0.45\textwidth]{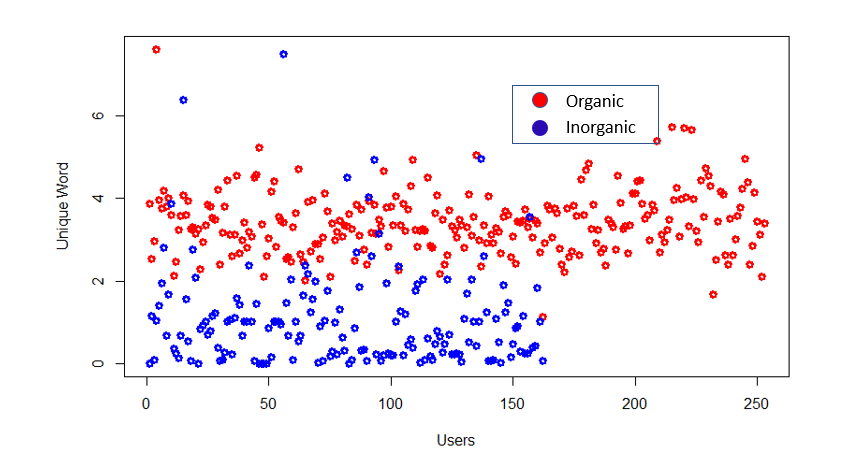}
    % }
    \caption{The left panel shows the Periodicity of Tweet Activity of Organic and Inorganic accounts, while the right panel shows the Number of Unique Words used by Organic and Inorganic Accounts}
    \label{fig2}
    \end{center}
    \end{figure}

    \item We employed an ARIMA (Autoregressive Integrated Moving Average) process to model the time series data and extract relevant features. These ARIMA-derived features included the log-likelihood, sum square of coefficients, error variance, and fit length. By incorporating these features, we aimed to capture the underlying temporal patterns and statistical characteristics of tweet activity, providing valuable insights for our classification model.
    % We also fitted the data to an ARIMA (Autoregressive Integrated Moving Average) process, and used the fit features in our classification model. In particular, the features taken from the ARIMA fit were: log-likelihood, sum square of coefficients, error variance, and fit length. 
    \item While periodicity identifies peak frequencies in the periodogram, we also considered the shape of the periodogram as an additional feature. To capture this aspect, we utilized local maxima in the periodogram, enabling us to characterize and incorporate the nuanced properties of tweet activity patterns beyond their primary periodicity.
    % Finally, we also included features to capture the shape of the periodogram. Periodicity only identifies the peak of the periodogram. However, we might also be interested in the shape of the periodogram, and hence we used local maxima to incorporate those properties as well.
\end{itemize}

% Apart the temporal features mentioned above, we also included semantic features from the tweet history as well. On inspection of the tweets, we found out that apart from the significant difference in the time series, there also appears to stark contrast in the words used by the organic and inorganic users. Inorganic users appear to use similar words repetitively, a trait that was expectantly absent in the tweet pattern of organic users. Following are the semantic features used in the classification algorithm:

In addition to the critical temporal features discussed previously, our classification algorithm also incorporated semantic features derived from the tweet histories of users. Upon closer examination of the tweets, a discernible disparity emerged in the linguistic patterns employed by organic and inorganic users. Inorganic users exhibited a notable propensity for repetitive usage of similar words—a characteristic conspicuously absent in the tweet patterns of organic users. The semantic features included in our classification algorithm to capture these distinctions are as follows:

\begin{itemize}
    \item \textbf{Average Number of Words per Tweet and Variance:} We calculated the average number of words per tweet and its variance for each user, providing insights into their linguistic behavior and variability in tweet length.
    \item \textbf{Number of Unique Words Used:} This feature quantifies the diversity of vocabulary employed by users, allowing us to differentiate between users who exhibit repetitive word usage and those with a more varied lexicon (see right panel of Figure \ref{fig2}).
    \item \textbf{Relative Frequency of Most Used Words:} We analyzed the relative frequency of the most commonly used words in a user's tweets. Inorganic users tend to employ a limited set of words repeatedly, while organic users typically exhibit greater lexical diversity.
    \item \textbf{Relative Hashtag Frequency: }The frequency of hashtags used in tweets was examined to discern any patterns associated with specific types of users or content. This feature aids in distinguishing between organic and inorganic users based on their hashtag usage behavior.
    \item \textbf{Sentiment Score:} Sentiment analysis was performed using three dictionaries available in R: AFINN, BING, and NRC. This yielded a sentiment score for each user, providing insights into the emotional tone of their tweets. Variations in sentiment can be indicative of differences in user behavior and content.
\end{itemize}

% \begin{figure}[htbp]
    
% \centerline{\includegraphics[width = 0.45\textwidth]{uniqueWors.png}}
% \caption{Number of Unique Words used by Organic and Inorganic Accounts}
% \label{fig3}
% \end{figure}

The integration of these semantic features into our classification algorithm complements the temporal features, enabling a more holistic understanding of user behavior on social media platforms. By capturing linguistic and content-related distinctions between organic and inorganic users, these features enhance the robustness and accuracy of our bot detection model.

\section{Bot Identification and Validation}
\label{sec:results}

The features described in Section \ref{sec:methods} comprises a diverse set of 19 features, encompassing both temporal and semantic attributes.  However, the presence of potentially linearly related features can introduce multicollinearity, a complication in classification problems. To address this issue, we employed the Variance Inflation Factor (VIF) as a feature selection method. VIF helps identify and retain a subset of features that exhibit low multicollinearity, ensuring the robustness of our classification algorithm (\cite{black2010multivariate}). In particular, 11 features were selected for our next stage of the algorithm.
% However, one or more of these features might be linearly related, resulting in multicollinearity in the data, which is a nuisance in classification problem. Hence, VIF (Variance Inflation Factor) is used to select 11 of these features in our classification algorithm. 
We have used two main types of classification algorithm: K-means clustering and SVM 
% and logistic regression based clustering.
K-Means clustering is an unsupervised classification algorithm designed to partition data into clusters based on the provided features \cite{hastie2009elements}. In our analysis, K-Means clustering aids in uncovering hidden patterns within the dataset, enabling us to distinguish between different user groups. The SVM (or Support Vector Machine) a supervised algorithm, constructs a classifier using a designated training dataset, which can subsequently be employed to classify new data points (\cite{cortes1995support}). Its application in our research facilitates accurate classification by creating a discriminative boundary between bot and non-bot users. 
% Finally, the logistic regression based algorithm involves fitting a logistic regression model to our dataset, where the dependent variable corresponds to the labels of the training dataset, and the independent variables consist of the extracted features (\cite{hosmer2013applied}). Predictions on the test dataset yield scores ranging from 0 to 1, which are then clustered into multiple partitions to classify users based on their behavior.

K-means clustering is an \textit{unsupervised learning} algorithm that partitions a dataset into \( K \) distinct clusters by minimizing the within-cluster variance. Given a dataset \( \{x_1, x_2, \dots, x_n\} \), where each \( x_i \) is a \( d \)-dimensional feature vector, the goal is to assign each point to one of \( K \) clusters, represented by centroids \( \{c_1, c_2, \dots, c_K\} \). The objective function to minimize is:
\[
J = \sum_{i=1}^{K} \sum_{x \in C_i} ||x - c_i||^2
\]
where \( C_i \) represents the set of points assigned to cluster \( i \) and \( c_i \) is the centroid of cluster \( i \), computed as the mean of all points in the cluster. The algorithm iteratively updates cluster assignments and centroid positions until convergence, typically using the Lloyd’s algorithm, ensuring that data points are grouped based on similarity.

Support Vector Machines (SVM) is a \textit{supervised learning} algorithm used for binary classification by finding an optimal decision boundary (hyperplane) that maximizes the margin between two classes. Given a labeled dataset \( \{(x_i, y_i)\}_{i=1}^{n} \), where \( x_i \in \mathbb{R}^d \) are feature vectors and \( y_i \in \{-1,1\} \) are class labels, SVM seeks a hyperplane defined by:
\[
w^T x + b = 0
\]
where \( w \) is the weight vector and \( b \) is the bias term. The \textit{margin} is maximized by solving the following optimization problem:
\[
\min_{w, b} \frac{1}{2} ||w||^2 \quad \text{subject to} \quad y_i (w^T x_i + b) \geq 1, \quad \forall i.
\]

For non-linearly separable data, SVM uses kernel functions \( K(x_i, x_j) \) to project data into a higher-dimensional space where it becomes linearly separable. The result is a robust classification model that generalizes well to unseen data.

\begin{figure}[htbp]
\centerline{\includegraphics[width = 0.45\textwidth]{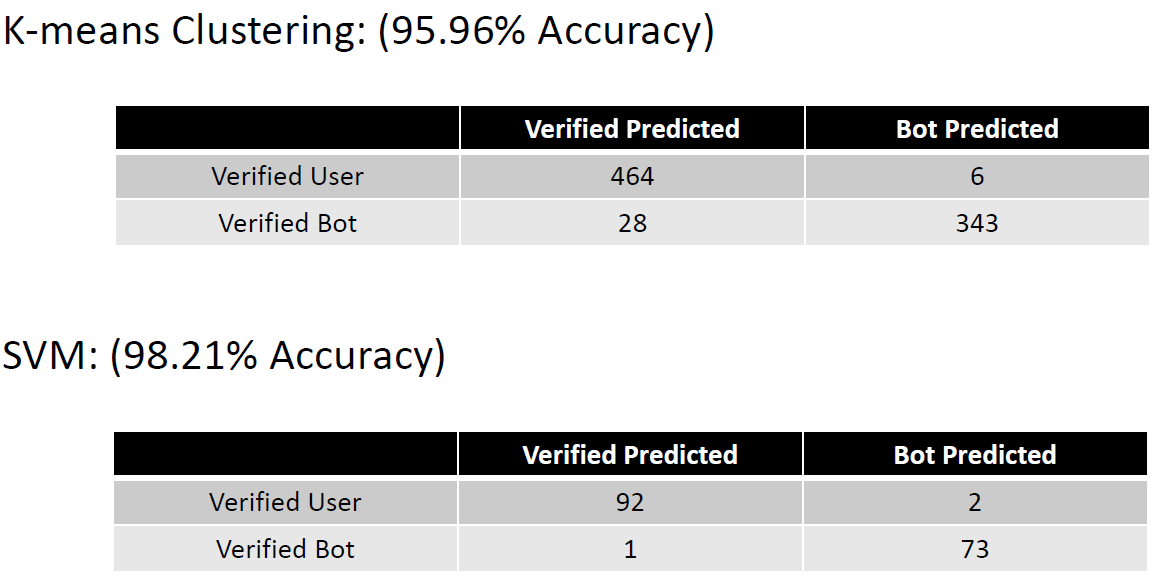}}
\caption{Clustering Accuracy of one unsupervised and one supervised machine learning algorithm}
\label{fig4}
\end{figure}

The output of the K-Means clustering and SVM classification algorithms is presented in Figure \ref{fig4}, providing a visual representation of the classification accuracy achieved by each method. The results of our classification experiments demonstrate notable levels of accuracy for all three employed algorithms. The K-means algorithm exhibits a \textbf{specificity}(True Negative Rate) of about $0.925$, \textbf{sensitivity} (True Positive Rate or Recall) of about $0.987$ and a \textbf{F-score} of about $0.965$. On the other hand, SVM demonstrated a \textbf{specificity} of about $0.986$, \textbf{sensitivity} of about $0.979$ and a \textbf{F-score} of about $0.984$. Specifically, both K-Means clustering and SVM exhibited commendable classification accuracy, with SVM outperforming K-Means by a slight margin. This outcome can be attributed to SVM's supervised nature, which allows it to learn from labeled training data. Furthermore, the Logistic Regression-based clustering approach yielded a particularly impressive accuracy rate of approximately 97\%. These findings underscore the effectiveness of our chosen methodologies and highlight the potential for robust bot detection in the context of social media data.

% As we can see, both k-means and SVM provided moderately well classification accuracy. SVM performs marginally better since it is a supervised machine learning algorithm, in contrast to k-means clustering. The Logistic regression based clustering also gave around 97\% accuracy.

With the attainment of noteworthy classification accuracy in our experiments, it becomes pertinent to investigate the significance of the features utilized within the clustering algorithm. The assessment of feature importance can be approached through various methodologies. In this study, we introduce the concept of an Accuracy Score (AS) for features, a metric designed to quantify the contribution of individual features to the overall classification accuracy. The Accuracy Score (AS) is computed using the following formula:
\begin{equation}
   AS = 100 \times \frac{Accuracy - Accuracy_i}{\sum_i \left(Accuracy - Accuracy_i\right)},
\end{equation}
where $Accuracy_i$ is the accuracy of the classification algorithm when the $i^{th}$ feature is removed. By calculating AS for each feature, we gain insights into the relative importance of individual features in driving classification accuracy. This analysis allows us to discern which features have the most substantial impact on the effectiveness of our clustering algorithm, further enriching our understanding of bot detection in social media datasets.

\begin{figure}[htbp]
\centerline{\includegraphics[width = 0.45\textwidth]{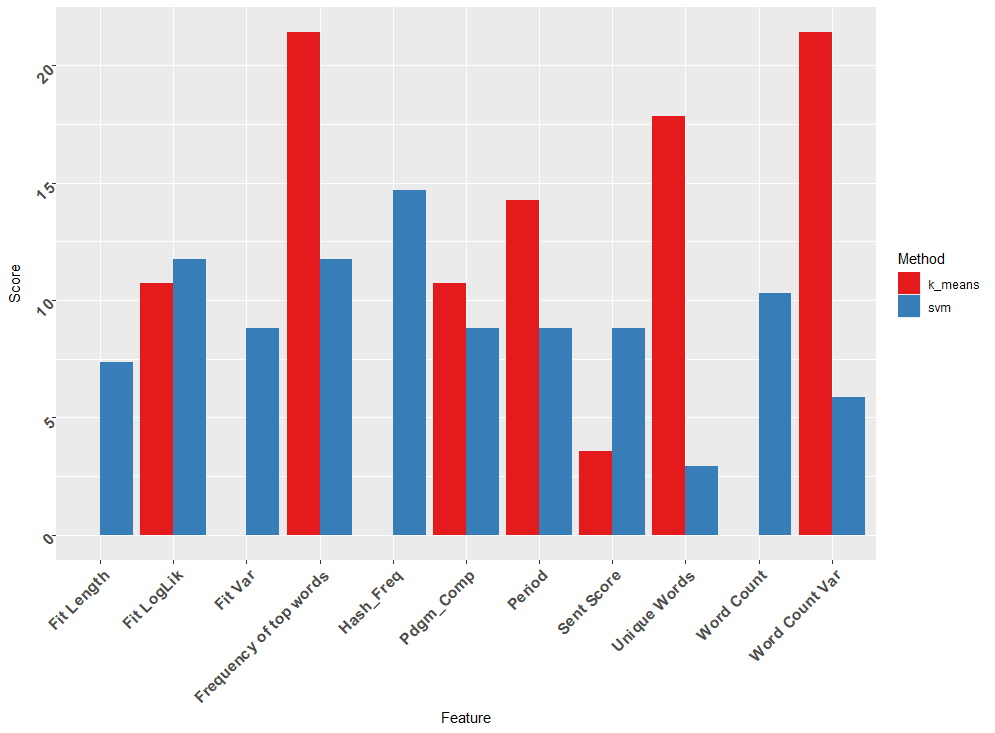}}
\caption{Accuracy Score of Features}
\label{fig6}
\end{figure}

Figure \ref{fig6} provides a comprehensive overview of the accuracy scores assigned to individual features under the influence of the two classification algorithms, K-Means, and SVM. Notably, the analysis reveals distinct patterns of feature importance for each algorithm. In the case of K-Means clustering, the relative frequency of words and word count variance emerge as the most influential features, signifying their pivotal role in the clustering process. These features likely capture critical patterns in user behavior that facilitate the effective separation of bot and non-bot users. Conversely, SVM classification prioritizes the importance of Hashtags and word count, underscoring their significance in distinguishing between organic and inorganic users. These features likely contribute to SVM's ability to create an optimal classification boundary in the feature space. 
% To further assess feature importance, we explored alternative machine learning techniques, such as Bagging and Boosting (\cite{sutton2005classification}), as well as LASSO (see Figure \ref{fig6}). These methods provided additional perspectives on feature relevance, shedding light on the nuanced contributions of different features to the classification process.

% Figure \ref{fig6} gives the accuracy score of the features under the two classification algorithms. As we can see, for k-means, relative frequency of words and word count variance are most significant features, while for SVM the Hashtag and word count are most important.

% We can also assess the importance of the features by other machine learning techniques, for example Bagging and Boosting, or LASSSO. Figure 4 gives the output for these two methods.

%  \begin{figure}[htbp]
% \centerline{\includegraphics[width = 0.45\textwidth]{Figures/imptnc.png}}
% \caption{LASSO and Bagging And Boosting}
% \label{fig6}
% \end{figure}

\section{Uncertainty Quantification}
\label{sec:uncertainty}

% \subsection{Conformal Prediction}

\subsection{Split Conformal Prediction}

Split Conformal Prediction (SCP), also known as \textit{Inductive Conformal Prediction (ICP)}, is a computationally efficient variation of Conformal Prediction that provides valid prediction sets with a pre-specified confidence level while avoiding the high computational cost associated with Full Conformal Prediction (FCP). The core idea behind SCP is to split the available data into a \textbf{training set} and a separate \textbf{calibration set}, where the latter is used to estimate p-values for uncertainty quantification. Given a dataset:
\[
\mathcal{D} = \{(X_1, Y_1), (X_2, Y_2), ..., (X_n, Y_n)\},
\]
SCP first partitions the data into two disjoint subsets: a \textbf{training set} \(\mathcal{D}_{\text{train}}\) and a \textbf{calibration set} \(\mathcal{D}_{\text{calib}}\). A predictive model \( f(X) \) is trained on \(\mathcal{D}_{\text{train}}\), and for each instance \((X_i, Y_i) \in \mathcal{D}_{\text{calib}}\), a \textbf{nonconformity score} is computed to measure how unusual a label is relative to the training data. A common choice for classification problems is:
\[
s(X_i, Y_i) = 1 - P(Y_i|X_i),
\]
where \( P(Y_i|X_i) \) represents the probability assigned to the correct class by the trained model \citep{lei2018distribution}.

For a new test instance \( X_{n+1} \), we calculate a p-value for each possible label \( y \) by comparing its nonconformity score with those from the calibration set:
\[
p(y) = \frac{ \sum_{i=1}^{|\mathcal{D}_{\text{calib}}|} \mathbf{1} \{ s(X_i, Y_i) \geq s(X_{n+1}, y) \} + 1 }{|\mathcal{D}_{\text{calib}}| + 1}.
\]
The final \textbf{prediction set} at confidence level \( 1 - \alpha \) is given by:
\[
\Gamma_{n+1}^{\alpha} = \{ y \in \mathcal{Y} \mid p(y) > \alpha \}.
\]
This ensures that, in expectation, the true label is contained in the prediction set with probability at least \( 1 - \alpha \), a property known as \textbf{marginal validity} \citep{vovk2005algorithmic}.

SCP has been widely adopted due to its \textbf{scalability and flexibility}. Unlike FCP, which requires retraining the model for each test instance, SCP only requires \textbf{one-time model training}, making it suitable for \textbf{high-dimensional datasets and real-time applications} \citep{lei2018distribution}. It has been successfully applied in various domains, including \textbf{medical diagnosis} \citep{angelopoulos2020uncertainty}, \textbf{natural language processing} \citep{fisch2021few}, and \textbf{financial risk assessment} \citep{papadopoulos2008inductive}. Furthermore, recent research has introduced \textbf{adaptive conformal prediction}, where calibration sets are dynamically adjusted to improve efficiency while preserving validity \citep{romano2019conformalized}. SCP remains a powerful tool for \textbf{uncertainty quantification}, providing rigorous statistical guarantees while maintaining computational feasibility in modern machine learning applications.

\subsection{Results}

\begin{figure}
    \centering
    \includegraphics[width=\linewidth]{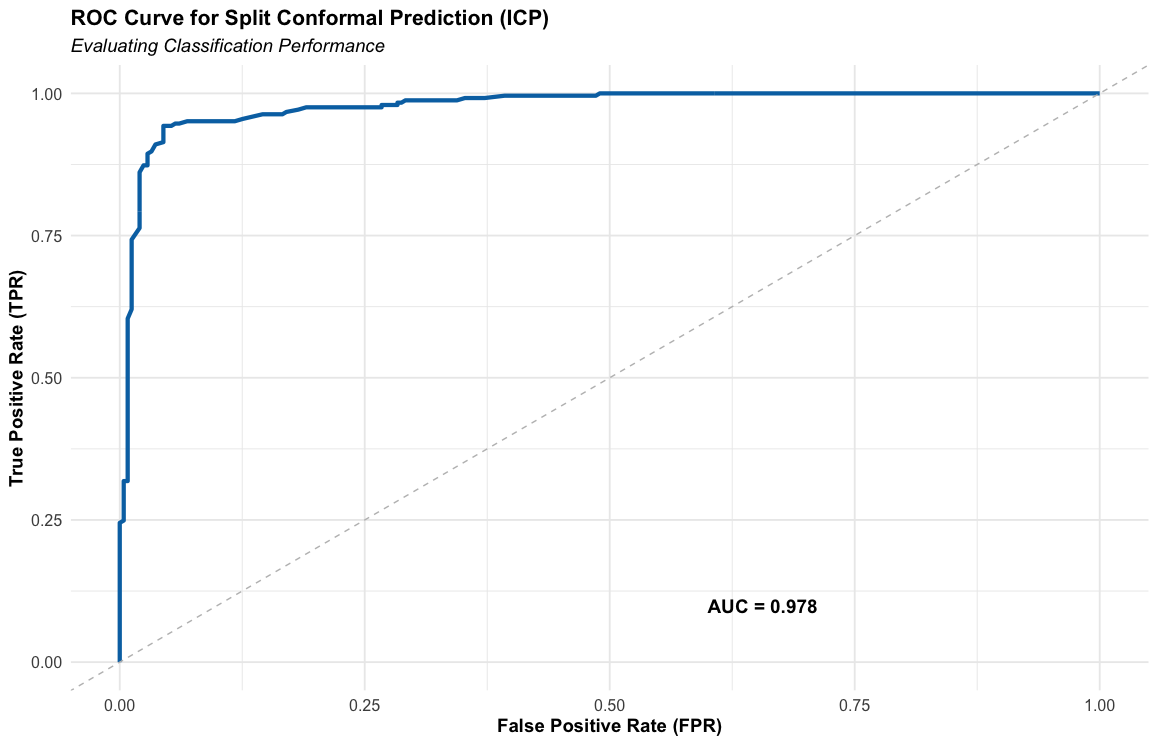}
    \caption{ROC Curve for Split Conformal Prediction}
    \label{fig:roc}
\end{figure}

The Split Conformal Prediction (ICP) method demonstrated strong classification performance while maintaining robust uncertainty quantification. The confusion matrix (Table \ref{tab:conf_matrix}) reveals that the model correctly classifies 234 negative instances and 232 positive instances, with only 13 misclassifications in each class, leading to a high overall accuracy of 94.72\%. The Kappa statistic of 0.8943 further confirms the model's strong agreement beyond chance. Additionally, the high sensitivity (0.9474) and specificity (0.9469) indicate that the model effectively differentiates between classes while minimizing false positives and false negatives. The positive and negative predictive values (PPV = 0.9474, NPV = 0.9469) confirm that the model maintains high reliability in both classifications. The McNemar’s test p-value of 1 suggests that there is no significant disagreement between misclassification rates, further validating the model's stability. The ROC curve analysis (Figure \ref{fig:roc}) reinforces these findings, demonstrating a high AUC of 0.978, indicating strong discrimination between classes. The steep increase in true positive rate (TPR) at low false positive rates (FPR) suggests that the model assigns high confidence to correct classifications. This aligns with the confusion matrix results, further confirming the model’s ability to produce well-calibrated confidence scores.
% The Split Conformal Prediction (ICP) method demonstrated strong classification performance while maintaining valid uncertainty quantification. The ROC curve analysis (Figure \ref{fig:roc}) revealed an AUC of 0.978, indicating that the model effectively differentiates between classes with minimal false positive rates. The curve exhibits a sharp increase near the origin, reflecting high sensitivity and specificity, confirming that the classifier consistently assigns high confidence scores to correct predictions. This suggests that the model provides well-calibrated probability estimates, which is crucial for reliable conformal inference.

\begin{figure}[h!]
    \centering
    \includegraphics[width=\linewidth]{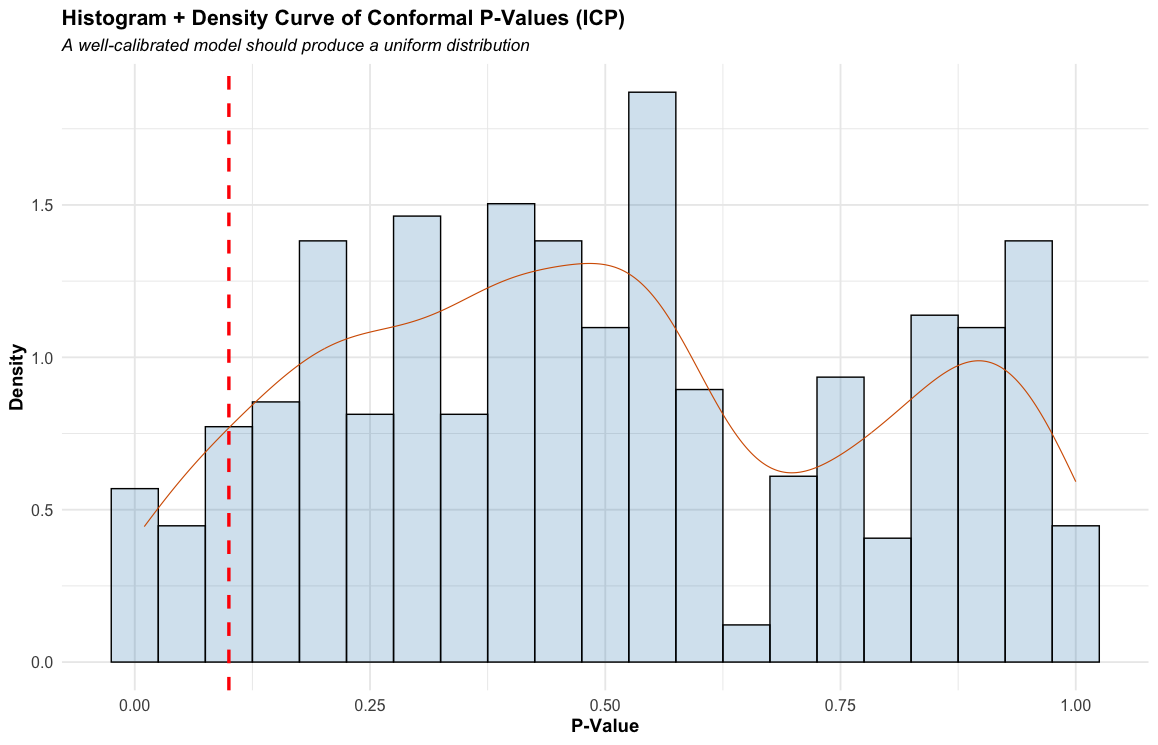}
    \caption{Histogram of Conformal P-Values for Split Conformal Prediction (ICP)}
    \label{fig:histP}
\end{figure}

Further evaluation of conformal p-values (Figure \ref{fig:histP}) provides insight into the model's calibration. A well-calibrated model should produce p-values that are uniformly distributed, ensuring that uncertainty is accurately captured. While the observed histogram exhibits approximate uniformity, minor fluctuations suggest some deviations in certain probability regions. The density curve overlays the histogram, further visualizing the p-value distribution, which remains relatively smooth, reinforcing the validity of the conformal prediction framework. The presence of a dashed red line at $\alpha = 0.1$ marks the decision threshold, ensuring that low-confidence predictions are correctly identified. 

\begin{table}[h]
    \centering
    \caption{Confusion Matrix and Performance Statistics for Split Conformal Prediction (ICP)}
    \label{tab:conf_matrix}
    \begin{tabular}{lcc}
        \toprule
        \textbf{Prediction vs. Reference} & \textbf{Actual: 0} & \textbf{Actual: 1} \\
        \midrule
        \textbf{Predicted: 0} & 234 & 13 \\
        \textbf{Predicted: 1} & 13  & 232 \\
        \bottomrule
    \end{tabular}

    \vspace{1em}

    \begin{tabular}{lc}
        \toprule
        \textbf{Metric} & \textbf{Value} \\
        \midrule
        Accuracy & 0.9472 \\
        95\% Confidence Interval & $(0.9235, 0.9652)$ \\
        No Information Rate & $0.502$ \\
        P-Value ($Acc > NIR$) & $< 2e-16$ \\
        Kappa Statistic & $0.8943$ \\
        McNemar's Test P-Value & $1$ \\
        Sensitivity & $0.9474$ \\
        Specificity & $0.9469$ \\
        Positive Predictive Value (PPV) & $0.9474$ \\
        Negative Predictive Value (NPV) & $0.9469$ \\
        Prevalence & $0.5020$ \\
        Detection Rate & $0.4756$ \\
        Detection Prevalence & $0.5020$ \\
        Balanced Accuracy & $0.9472$ \\
        \bottomrule
    \end{tabular}
    
    \vspace{1em}
    \textbf{Positive Class:} 0

\end{table}

\begin{figure}[h!]
    \centering
    \includegraphics[width=0.5\linewidth]{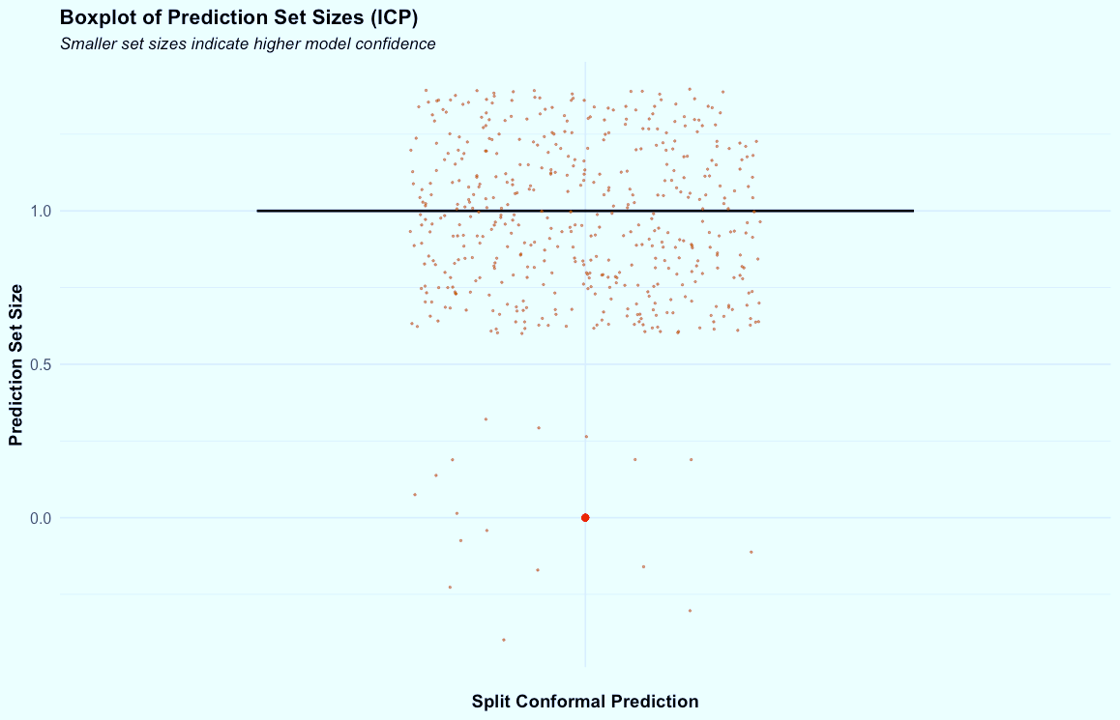}
    \caption{Boxplot of Prediction Set Sizes}
    \label{fig:boxOutlier}
\end{figure}

The boxplot of prediction set sizes (Figure \ref{fig:boxOutlier}) highlights the model’s ability to adapt to different levels of uncertainty. The majority of instances receive single-label predictions, confirming that the model is confident in most cases, while a subset of samples receives multi-label predictions, reflecting cases where uncertainty is higher. The jitter plot overlay provides additional clarity on the distribution of set sizes, showing that while most predictions are confident, the model still allows for conservative predictions when needed. The presence of occasional outliers, including instances with empty prediction sets, suggests that some samples may be assigned overconfident predictions, which could be further addressed through additional calibration techniques.

Overall, the results confirm that Split Conformal Prediction offers a balance between classification accuracy and uncertainty quantification, ensuring that high-confidence predictions remain precise while maintaining a principled approach to handling uncertainty. The model’s strong classification performance and ability to provide well-calibrated prediction sets make it an effective and computationally efficient choice for conformal inference.

\section{Conclusion}
\label{sec:conslusion}

% In this study, we have explored static time bot detection, a method that leverages historical tweet data to classify users as either bots or non-bots. Our investigation has demonstrated that through the application of straightforward machine learning techniques, we can achieve highly accurate classification results, characterized by elevated levels of specificity and sensitivity. Additionally, we have provided a feature ranking based on importance, shedding light on the key determinants behind the success of these classification algorithms. Static bot detection, while renowned for its reliability, presents certain limitations. It demands an extensive corpus of tweets from each user and is contingent on a user's prior tweet activity, rendering it time-consuming and potentially restrictive in its applicability. As a result, future research endeavors may be inclined to explore the development of dynamic time bot-detection techniques. Such approaches would offer the advantage of providing real-time scoring and adaptability to evolving user behavior, potentially addressing the limitations of static bot detection. In summary, our findings underscore the efficacy of static time bot detection while recognizing the need for innovative methodologies that can adapt to the dynamic nature of online interactions. These insights contribute to the ongoing discourse on enhancing bot detection in the context of social media, with implications for bolstering online security and trustworthiness.

In this study, we presented a machine learning-based framework for bot detection in social media by leveraging semantic, behavioral, and temporal features extracted from user-generated content. By employing Support Vector Machines (SVM) alongside k-means clustering, our approach successfully distinguished between human and automated accounts, demonstrating high classification accuracy, specificity, and sensitivity. The ROC analysis confirmed the strong discriminative power of the model, with a high AUC value, indicating that the chosen feature set effectively captures the characteristics that differentiate bots from human users. Additionally, the analysis of conformal p-values provided valuable insights into the model’s uncertainty estimation, confirming its reliability in classification tasks while maintaining valid prediction sets.

The study also explored the calibration of conformal prediction sets, which ensured that the classifier not only produced accurate predictions but also provided well-calibrated confidence scores. The histogram of conformal p-values exhibited approximate uniformity, indicating proper calibration, while the boxplot of prediction set sizes highlighted the model’s ability to adjust its confidence dynamically. Instances receiving single-label predictions reflected high confidence, whereas multi-label predictions were observed for cases where uncertainty was higher, ensuring that ambiguous cases were handled appropriately. Furthermore, the confusion matrix analysis demonstrated strong model performance, with low misclassification rates, a Kappa statistic of 0.8943, and balanced sensitivity and specificity, reinforcing the robustness of the proposed method.

While our approach achieves high detection accuracy and strong uncertainty quantification, certain limitations must be addressed to enhance its scalability and adaptability to evolving bot behaviors. First, the method primarily relies on historical tweet data, making it less adaptive to real-time bot evolution. Future work should focus on incorporating dynamic, real-time detection models, which leverage adaptive learning techniques to recognize emerging manipulation strategies in social media. Additionally, integrating graph-based and network analysis methods could further strengthen the model’s ability to detect coordinated bot activities and adversarial networks. Another promising direction involves deep learning architectures, such as transformer-based models, which can process large-scale text data more effectively while preserving contextual nuances.

Beyond its technical contributions, this study has practical implications for social media security, misinformation detection, and platform integrity enforcement. The proposed framework provides a scalable and interpretable solution for detecting automated accounts, which can be integrated into platform moderation systems, cybersecurity infrastructures, and misinformation tracking pipelines. As bots continue to evolve and adopt more sophisticated evasion techniques, it is imperative to develop adaptive, explainable, and robust detection frameworks that ensure the credibility and trustworthiness of online ecosystems. Through continued advancements in machine learning, network science, and adversarial bot detection, we can build resilient systems to combat automated disinformation and enhance digital platform security.

\bibliography{ref}

\begin{thebibliography}{24}
\providecommand{\natexlab}[1]{#1}
\providecommand{\url}[1]{\texttt{#1}}
\expandafter\ifx\csname urlstyle\endcsname\relax
  \providecommand{\doi}[1]{doi: #1}\else
  \providecommand{\doi}{doi: \begingroup \urlstyle{rm}\Url}\fi

\bibitem[Angelopoulos et~al.(2020)Angelopoulos, Bates, Malik, and Jordan]{angelopoulos2020uncertainty}
Anastasios Angelopoulos, Stephen Bates, Jitendra Malik, and Michael~I Jordan.
\newblock Uncertainty sets for image classifiers using conformal prediction.
\newblock \emph{arXiv preprint arXiv:2009.14193}, 2020.

\bibitem[Bessi and Ferrara(2016)]{bessi2016social}
Alessandro Bessi and Emilio Ferrara.
\newblock Social bots distort the 2016 us presidential election online discussion.
\newblock \emph{First monday}, 21\penalty0 (11-7), 2016.

\bibitem[Black et~al.(2010)Black, Babin, and Anderson]{black2010multivariate}
William~C Black, Barry~J Babin, and Rolph~E Anderson.
\newblock \emph{Multivariate data analysis: A global perspective}.
\newblock Pearson, 2010.

\bibitem[Cortes and Vapnik(1995)]{cortes1995support}
Corinna Cortes and Vladimir Vapnik.
\newblock Support-vector networks.
\newblock \emph{Machine learning}, 20:\penalty0 273--297, 1995.

\bibitem[Cresci et~al.(2015)Cresci, Di~Pietro, Petrocchi, Spognardi, and Tesconi]{cresci2015fame}
Stefano Cresci, Roberto Di~Pietro, Marinella Petrocchi, Angelo Spognardi, and Maurizio Tesconi.
\newblock Fame for sale: Efficient detection of fake twitter followers.
\newblock \emph{Decision Support Systems}, 80:\penalty0 56--71, 2015.

\bibitem[Cresci et~al.(2017)Cresci, Di~Pietro, Petrocchi, Spognardi, and Tesconi]{cresci2017social}
Stefano Cresci, Roberto Di~Pietro, Marinella Petrocchi, Angelo Spognardi, and Maurizio Tesconi.
\newblock Social fingerprinting: detection of spambot groups through dna-inspired behavioral modeling.
\newblock \emph{IEEE Transactions on Dependable and Secure Computing}, 15\penalty0 (4):\penalty0 561--576, 2017.

\bibitem[Davis et~al.(2016)Davis, Varol, Ferrara, Flammini, and Menczer]{davis2016botornot}
Clayton~Allen Davis, Onur Varol, Emilio Ferrara, Alessandro Flammini, and Filippo Menczer.
\newblock Botornot: A system to evaluate social bots.
\newblock In \emph{Proceedings of the 25th international conference companion on world wide web}, pages 273--274, 2016.

\bibitem[Duggan et~al.(2013)Duggan, Brenner, et~al.]{duggan2013demographics}
Maeve Duggan, Joanna Brenner, et~al.
\newblock \emph{The demographics of social media users, 2012}, volume~14.
\newblock Pew Research Center's Internet \& American Life Project Washington, DC, 2013.

\bibitem[Ferrara et~al.(2016)Ferrara, Varol, Davis, Menczer, and Flammini]{ferrara2016rise}
Emilio Ferrara, Onur Varol, Clayton Davis, Filippo Menczer, and Alessandro Flammini.
\newblock The rise of social bots.
\newblock \emph{Communications of the ACM}, 59\penalty0 (7):\penalty0 96--104, 2016.

\bibitem[Fisch et~al.(2021)Fisch, Schuster, Jaakkola, and Barzilay]{fisch2021few}
Adam Fisch, Tal Schuster, Tommi Jaakkola, and Regina Barzilay.
\newblock Few-shot conformal prediction with auxiliary tasks.
\newblock In \emph{International Conference on Machine Learning}, pages 3329--3339. PMLR, 2021.

\bibitem[Ghosh et~al.(2025)Ghosh, Boettcher, Johnston, and Lahiri]{ghosh2025thanos}
Dhrubajyoti Ghosh, William~A Boettcher, Rob Johnston, and Soumendra Lahiri.
\newblock {THANOS}: A predictive model of electoral campaigns using twitter data and opinion polls.
\newblock \emph{Data Science in Science}, 4\penalty0 (1):\penalty0 2484180, 2025.

\bibitem[Hastie et~al.(2009)Hastie, Tibshirani, Friedman, and Friedman]{hastie2009elements}
Trevor Hastie, Robert Tibshirani, Jerome~H Friedman, and Jerome~H Friedman.
\newblock \emph{The elements of statistical learning: data mining, inference, and prediction}, volume~2.
\newblock Springer, 2009.

\bibitem[Lee et~al.(2011)Lee, Eoff, and Caverlee]{lee2011seven}
Kyumin Lee, Brian Eoff, and James Caverlee.
\newblock Seven months with the devils: A long-term study of content polluters on twitter.
\newblock In \emph{Proceedings of the international AAAI conference on web and social media}, volume~5, pages 185--192, 2011.

\bibitem[Lei et~al.(2018)Lei, G’Sell, Rinaldo, Tibshirani, and Wasserman]{lei2018distribution}
Jing Lei, Max G’Sell, Alessandro Rinaldo, Ryan~J Tibshirani, and Larry Wasserman.
\newblock Distribution-free predictive inference for regression.
\newblock \emph{Journal of the American Statistical Association}, 113\penalty0 (523):\penalty0 1094--1111, 2018.

\bibitem[Mazza et~al.(2019)Mazza, Cresci, Avvenuti, Quattrociocchi, and Tesconi]{mazza2019rtbust}
Michele Mazza, Stefano Cresci, Marco Avvenuti, Walter Quattrociocchi, and Maurizio Tesconi.
\newblock Rtbust: Exploiting temporal patterns for botnet detection on twitter.
\newblock In \emph{Proceedings of the 10th ACM conference on web science}, pages 183--192, 2019.

\bibitem[Papadopoulos(2008)]{papadopoulos2008inductive}
Harris Papadopoulos.
\newblock Inductive conformal prediction: Theory and application to neural networks.
\newblock In \emph{Tools in artificial intelligence}. Citeseer, 2008.

\bibitem[Pennycook et~al.(2020)Pennycook, Bear, Collins, and Rand]{pennycook2020implied}
Gordon Pennycook, Adam Bear, Evan~T Collins, and David~G Rand.
\newblock The implied truth effect: Attaching warnings to a subset of fake news headlines increases perceived accuracy of headlines without warnings.
\newblock \emph{Management science}, 66\penalty0 (11):\penalty0 4944--4957, 2020.

\bibitem[Pew(2021)]{pew}
Pew.
\newblock Social media fact sheet, 2021.
\newblock URL \url{https://www.pewresearch.org/internet/fact-sheet/social-media/}.

\bibitem[Romano et~al.(2019)Romano, Patterson, and Candes]{romano2019conformalized}
Yaniv Romano, Evan Patterson, and Emmanuel Candes.
\newblock Conformalized quantile regression.
\newblock \emph{Advances in neural information processing systems}, 32, 2019.

\bibitem[Sayyadiharikandeh et~al.(2020)Sayyadiharikandeh, Varol, Yang, Flammini, and Menczer]{sayyadiharikandeh2020detection}
Mohsen Sayyadiharikandeh, Onur Varol, Kai-Cheng Yang, Alessandro Flammini, and Filippo Menczer.
\newblock Detection of novel social bots by ensembles of specialized classifiers.
\newblock In \emph{Proceedings of the 29th ACM international conference on information \& knowledge management}, pages 2725--2732, 2020.

\bibitem[Subrahmanian et~al.(2016)Subrahmanian, Azaria, Durst, Kagan, Galstyan, Lerman, Zhu, Ferrara, Flammini, and Menczer]{subrahmanian2016darpa}
Venkatramanan~S Subrahmanian, Amos Azaria, Skylar Durst, Vadim Kagan, Aram Galstyan, Kristina Lerman, Linhong Zhu, Emilio Ferrara, Alessandro Flammini, and Filippo Menczer.
\newblock The darpa twitter bot challenge.
\newblock \emph{Computer}, 49\penalty0 (6):\penalty0 38--46, 2016.

\bibitem[Tanwar et~al.(2021)Tanwar, Chaudhry, and Srivastava]{tanwar2021influencer}
Anshika~Singh Tanwar, Harish Chaudhry, and Manish~Kumar Srivastava.
\newblock Influencer marketing as a tool of digital consumer engagement: A systematic literature review.
\newblock \emph{Indian Journal of Marketing}, 51\penalty0 (10):\penalty0 27--42, 2021.

\bibitem[Tufekci(2017)]{tufekci2017twitter}
Zeynep Tufekci.
\newblock \emph{Twitter and tear gas: The power and fragility of networked protest}.
\newblock Yale University Press, 2017.

\bibitem[Vovk et~al.(2005)Vovk, Gammerman, and Shafer]{vovk2005algorithmic}
Vladimir Vovk, Alexander Gammerman, and Glenn Shafer.
\newblock \emph{Algorithmic learning in a random world}, volume~29.
\newblock Springer, 2005.

\end{thebibliography}
\bibliographystyle{plainnat}
% \bibliographystyle{IEEEtranN}

% \newpage

 

% \end{IEEEbiography}



% \vfill

\end{document}